\documentclass[12pt]{iopart}
\usepackage{graphicx}
\usepackage[dvips]{color}
\usepackage{pifont}
\usepackage{amssymb}

\newcommand{\be}{\begin{equation}}
\newcommand{\ee}{\end{equation}}
\newcommand{\bea}{\begin{eqnarray}}
\newcommand{\eea}{\end{eqnarray}}

\def\NL{\nonumber\\}
\def\eq#1{Eq.~(\ref{#1})}

\def\ds{\displaystyle}  \def\sss{\scriptstyle}

\def\tr#1{{\rm tr}\left\{#1\right\}}
\def\mm#1#2{{\renewcommand{\arraycolsep}{0.2em}%
\left(\begin{array}{#1}#2\end{array}\right)}}

\def\sx{\sigma^x} \def\sy{\sigma^y} \def\sz{\sigma^z}
\def\bv{b^{\rule{0mm}{2mm}}}
\def\bh{b^{\dagger}}

\def\gv{\gamma^{\rule{0mm}{2mm}}}
\def\ge{\gamma^{*}}
\def\lv{\lambda^{\rule{0mm}{2mm}}}
\def\le{\lambda^*}
\def\lvv{{\vec \lambda}^{\rule{0mm}{2mm}}}
\def\lve{{\vec \lambda}^*}

\def\mvv{{\vec \mu}^{\rule{0mm}{2mm}}}
\def\mve{{\vec \mu}^*}
\def\nvv{{\vec \nu}^{\rule{0mm}{2mm}}}
\def\nve{{\vec \nu}^*}
\def\gvv{{\vec \gamma}^{\rule{0mm}{2mm}}}
\def\gve{{\vec \gamma}^*}
\def\rvv{{\vec \rho}^{\rule{0mm}{2mm}}}
\def\rve{{\vec \rho}^*}
 
\newcommand{\mO}{\mathcal{O}}  

\newcommand{\suet}[1]{$\mathfrak{#1}$}
\newcommand{\sM}{\mbox{\suet{M}}} 
\newcommand{\suz}{\mbox{\suet{Z}}}

\newcommand{\Ttnull}{{\tilde T}_0}
\newcommand{\Tteins}{{\tilde T}_1}
\newcommand{\Ttzwei}{{\tilde T}_2}

\begin{document}

\title{Correlations in the Ising antiferromagnet on the anisotropic kagome lattice}

\author{Walter Apel$^1$ and Hans-Ulrich Everts$^2$}

\address{$^1$ Physikalisch-Technische Bundesanstalt (PTB), 
Bundesallee 100, 38116 Braunschweig, Germany}
\address{$^2$ Institut f\"ur Theoretische Physik, Leibniz Universit\"at Hannover,
Appelstra\ss e 2, 30167 Hannover, Germany} 
\ead{Walter.Apel@ptb.de and everts@itp.uni-hannover.de}

\begin{abstract}
We study the correlation function of  middle spins, i.~e.~of spins on 
intermediate sites between two adjacent parallel lattice axes, of the 
spatially anisotropic Ising antiferromagnet on the kagome lattice. 
It is given rigorously by a Toeplitz determinant.
The large-distance behaviour of this correlation function is obtained 
by analytic methods. 
For shorter distances we evaluate the Toeplitz determinant numerically. 
The correlation function is found to vanish exactly on a line $J_d(T)$ in 
the $T-J$ (temperature vs.~coupling constant)
 phase diagram. 
This disorder line divides the phase diagram into two regions. 
For $J < J_d(T)$ the correlations display  the features of an unfrustrated 
two-dimensional Ising magnet, whereas for $J > J_d(T)$ the correlations 
between the middle spins are seen to be  strongly influenced by the short-range
antiferromagnetic order that prevails among the spins of the adjacent lattice 
axes. 
While for  $J < J_d(T)$ there is a region with ferrimagnetic long-range order, 
the model remains disordered for  $J > J_d(T)$ down to $T=0$.            
\end{abstract}
\submitto{{\it JSTAT}}
\maketitle

\section{Introduction \label{intro} }

\begin{figure}
\parbox{0.45\linewidth}{
\includegraphics[height=6cm]{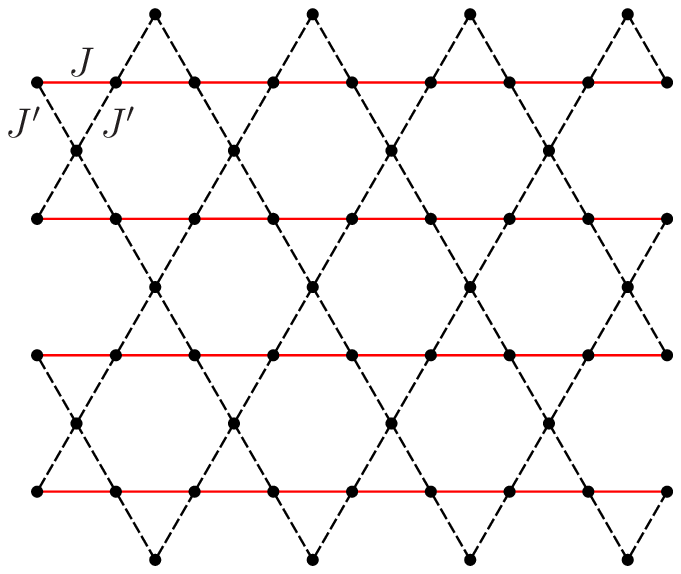}}\hspace{5mm}
\parbox{0.49\linewidth}{
\includegraphics[height=6cm]{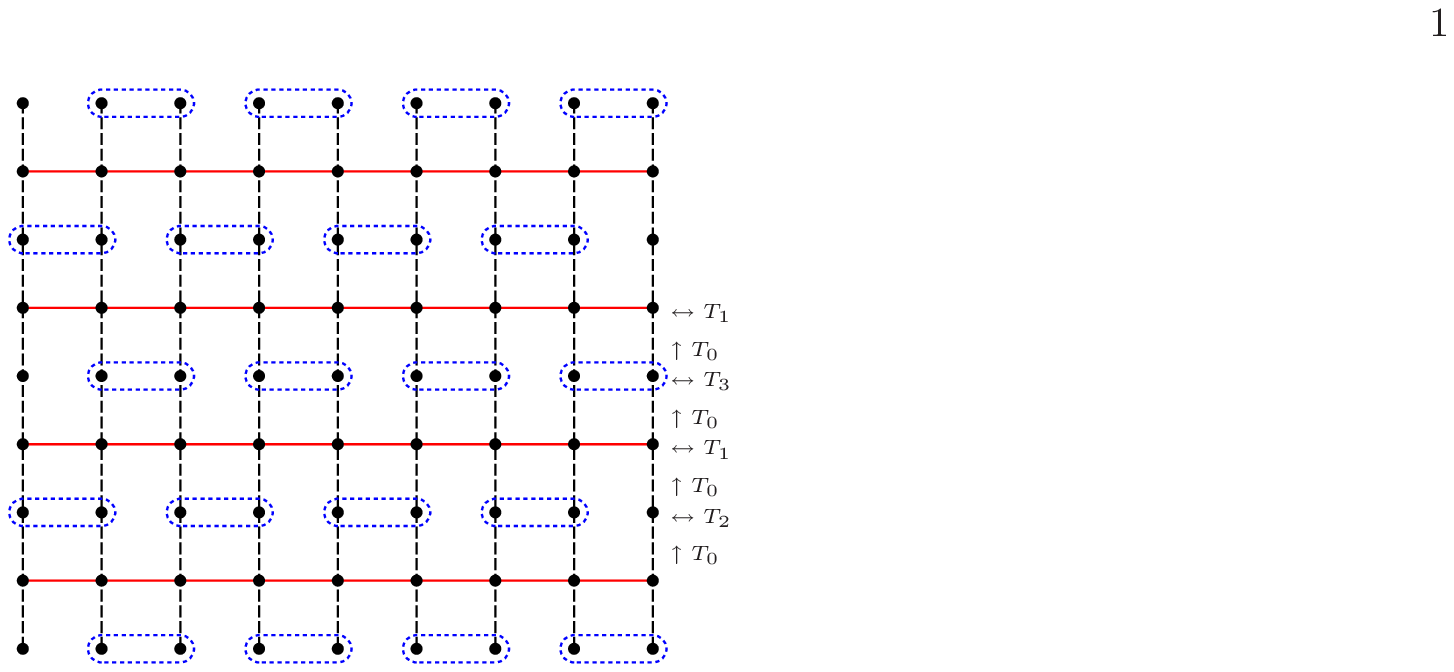}}
\caption{\label{aniskago} Anisotropic kagome model (left) and auxiliary model (right). 
The factors of the transfer matrix acting in and between the different rows 
are indicated on the right of the auxiliary model. }
\end{figure}

Recent experimental work on the mineral volborthite has stimulated several theoretical 
investigations of the Heisenberg antiferromgnet on the spatially anisotropic kagome lattice, 
 figure \ref{aniskago}, 
\cite{YAE07, WVK07,SSB08}. In this model, the exchange coupling in one of the three lattice
 directions 
($J$) differs from the couplings in the two other directions ($ J^\prime$).

The spin ensemble is thus 
divided into a set of parallel chains of spins with coupling $J$ and a set of middle 
spins, which are coupled to 
the chain spins  by the exchange constant $J'$. The results for the ground state of this model, 
obtained in the studies \cite{YAE07, WVK07,SSB08}, differ substantially from each other, particularly 
in the quasi-one-dimensional limit $J \gg J^\prime $ where the spin chains are only weakly 
coupled to 
the middle spins .\\
In this article, motivated by the difficulties encountered in approximate treatments
of the Heisenberg model, we address the Ising model with anisotropic couplings on the kagome 
lattice (AIK).
As was first demonstrated by Kano and Naya \cite{KN53}, this model can be solved exactly.
Its thermodynamic properties are discussed in detail in  recent work by Wei Li et~al. 
\cite{WSYSSG10}. For $J \neq J'$, the ground states are  
 immediately clear. In the region $J < J'$,  the ground state is
 ferrimagnetically ordered,
i.~e.~the spins on the chains are ordered ferromagnetically, and so are the middle spins,
 but their direction is antiparallel to the chain spins. For $J > J'$, the chain spins 
are ideally 
 ordered antiferromagnetically in the ground state leaving the middle
spins completely uncoupled. As a consequence, the chains are also decoupled from one 
another, so that the spins of each chain can form either one of the two antiferromagnetic orders.  
A simple counting argument yields a rest entropy of  
$\mathcal{S}^{rest}_{aniso}= \frac{1}{3} \ln2 \simeq 0.231 $ per spin in this case.
For the isotropic model, $J = J'$, the value of the rest entropy does not follow from 
a simple counting argument.  Kano and Naya \cite{KN53}  find in this case 
\newpage
\begin{eqnarray}
\mathcal{S}^{rest}_{iso} &=& \frac{1}{24 \pi^2} \int^{2\pi}_0\! \int^{2\pi}_0 \ln \big[ 21 - 
 4 ( \cos(\omega_1) +  \cos(\omega_2) +  \cos(\omega_1 + \omega_2)) \big] d\omega_1 d\omega_2 \NL[3mm]
&\simeq & 0.5018\;. 
\label{Srest}
\end{eqnarray}
 
In the present paper, we shall reformulate the transfer matrix formalism of Kano and Naya 
with the aim to  obtain not only the partition function, but also the 
correlation function $\chi(m,n)$ between two middle spins in 
the same row (see \fref{aniskago}). We expect this to be the most interesting 
of the various 
spin-spin correlation functions of the AIK model.\\ 
 The article is organised as follows. 
In section 2, we consider a slightly modified model that has been devised by Kano and Naya to 
allow for the application of the transfer matrix method, and that reproduces the AIK model in a 
certain limit. We present the transfer matrix, discuss its symmetry properties and convert 
it into a product of Grassmann functionals \cite{A86}. In section 3, we
 discuss the structure 
of the eigenvectors of the transfer matrix and calculate its eigenvalues. The latter are then 
used to construct the phase diagram of the AIK model as far as this is possible without 
knowledge of the correlation function. Section 4 is the central part of this article; it is 
devoted to our study of the above mentioned
 two-spin correlation. As in numerous 
previous investigations of correlations in Ising models on various two-dimensional lattices 
 \cite{MPW62, ST64, SML64, W66, ST70, AM74, G80, WZ82a, WZ82b}, the correlation function 
considered here is found to be represented rigorously by 
a Toeplitz determinant. 
By evaluating this determinant \cite{McCoyWu,Grenander-Szegoe}
we obtain analytic expressions for $\chi(m,n)$ in the limit of large 
distances between the spins, $|n - m| \gg 1$,  in various regions and on various 
special lines of the phase diagram. 
We support 
our asymptotic results by exact numerical evaluations of the Toeplitz determinant, which yield 
the correlation function also for short distances. In section 5 we present and 
discuss our results. 
Technicalities of the calculations that lead to these results are contained in appendices A, B, C.
In appendix D, we present a simple perturbative calculation, which is valid in the limit of 
large chain coupling, $J/J'\gg 1$, and which qualitatively reproduces the behaviour of the 
correlation function in this limit.
\\
After completion of this work, we became aware of a publication by M.~B.~Geilikman \cite{G74},
which addresses the same topic as the present paper. Using the methods of Vdovichenko 
\cite{V65a,V65b}, 
Geilikman arrives at the same Toeplitz determinant as  we, to represent the spin-spin 
correlation function. However, his evaluation of this determinant yields results that differ 
qualitatively from our expressions for $\chi(m,n)$ in the most interesting part of the phase 
diagram.

\section{Transfer matrices}

Following Kano and Naya \cite{KN53}, we split each intermediate spin of the AIK into two
Ising variables $\sigma_j, \sigma_{j+1}$, which are coupled ferromagnetically with a coupling $l$. 
At some point of our calculation we take the limit $ l \rightarrow \infty$ so that these two 
Ising spins freeze into the original intermediate spin. The auxiliary model, which results from 
the splitting, 
is shown in the right part of \fref{aniskago}. Obviously, it is periodic with period 4 in the 
(vertical) transfer direction. 
We consider a lattice of $M$ ($M$ divisible by 4) rows with $N$ 
($N$ even) sites in each row and impose periodic boundary conditions in both lattice directions. 
For the auxiliary model, the transfer matrix, whose highest eigenvalue yields the free energy 
of this model, is given by (see \fref{aniskago})

\be
T_0\;T_2\; T_0\; T_1\; T_0\; T_3\; T_0\; T_1,
\label{tf}
\ee

where
\bea
T_0 &=& (e^{2K'} - e^{-2K'})^{N/2} \; \prod_{n=0}^{N-1} \sx_n \;\;\; 
 e^{ a \sum_{n=0}^{N-1} \sx_n }, \label{tf0} \\
T_1 &=& e^{ -K \sum_{n=0}^{N-1} \sz_n \, \sz_{n+1}}, \label{tf1} \\
T_2 &=& e^{ \ell \sum_{n=0}^{(N-2)/2}(\sz_{2n+1} \, \sz_{2n+2}-1)}, \label{tf2} \\
T_3 &=& e^{ \ell \sum_{n=0}^{(N-2)/2}(\sz_{2n} \, \sz_{2n+1}-1)}, \label{tf3}
\eea 

with
\be
K = J/T \quad, \quad K^\prime = J^\prime/ T \quad \mbox{and} \quad \tanh(a)=e^{-2K^\prime}.
\label{KKa} 
\ee 
(Boltzmann's constant $k_B$ is set equal to unity throughout this paper.)\\
Instead of $K$ and $K'$ we shall use 

\be
\Psi= \tanh(K)\quad \mbox{and}\quad \Phi=\tanh(K') 
\label{phipsi}
\ee
as variables in the sequel.

The expression (\ref{tf}) is only one of various possibilities to represent the transfer matrix
of the auxiliary model. In fact, it is an inconvenient representation, since it is not symmetric.
Instead we can work with the following matrix    

\be
 T_2^{1/2}\; T_0\; T_1\; T_0\; T_3^{1/2}\;\cdot\; T_3^{1/2}\; T_0\; T_1\; T_0\; T_2^{1/2},\NL
\ee

which in turn is equivalent to

\be
T =  T_2\; T_0\; T_1\; T_0\; T_3\;\cdot\; T_3\; T_0\; T_1\; T_0\; T_2 \label{stf} 
\ee

in the limit $l \to \infty$, since $T_2$ and  $T_3$ become projectors in this limit, 
$T_i^2 = T_i,\; i=2,3$. 
$T$ is real and symmetric
since $T_3\;=\;s\;T_2\;s$, where  $s$ shifts  all spin 
variables by one lattice unit, $s\;\sx_n\;s =\sx_{n+1},\;\; s\;\sz_n\;s = \sz_{n+1}$, and 
therefore commutes with $T_0$ and $T_1$.

In the sequel, we follow the transfer-matrix method of 
Schulz, Mattis and Lieb \cite{SML64}. The necessary calculations are straight-forward. 
However, since the kagome lattice and the auxiliary lattice are non-Bravais lattices, some 
pecularities arise in comparison to the square lattice case. Therefore, we sketch the 
transfer-matrix procedure briefly here. 

In order to exhibit the free fermion nature of the AIK model,
one transforms the expressions in the exponents of
$T_0,\;...\;T_3$ into quadratic forms in fermion operators $b_n,\;b^{\dagger}_n$ by applying the 
Jordan-Wigner transform \cite{JW28}:
\bea
 \sx_n &=& 2 \bh_n \bv_n -1, \label{sx}\\
\sz_0 &=& b^{\dagger}_0 + b_0 \label{sz0} \quad \mbox{and} \quad 
 \sz_n = (\bh_n + \bv_n ) \;\; e^{i\pi \sum_{j=0}^{n-1} \bh_j \bv_j },\;\;0 < n <N
 \label{szn}.
\eea
This yields
\bea
 T_0 &=&  \left(\frac{ 4\Phi}{1-\Phi^2}\right)^{N/2} \; e^{ -Na}
 \; \; e^{(2a+i\pi) \sum_{n=0}^{N-1}  \bh_n \bv_n  },\label{T0}\\
 T_1 &=&   e^{ -K \sum_{n=0}^{N-1}(\bh_n-\bv_n)(\bh_{n+1}+\bv_{n+1})},\label{T1}\\
 T_2 &=&   e^{ \ell \sum_{n=0}^{(N-2)/2}
    \left[(\bh_{2n+1}-\bv_{2n+1})(\bh_{2n+2}+\bv_{2n+2}) -1\right]}, \label{T2} \\
 T_3 &=&   e^{\ell \sum_{n=0}^{(N-2)/2}
    \left[(\bh_{2n}-\bv_{2n})(\bh_{2n+1}+\bv_{2n+1}) -1\right]}, \label{T3}
\eea         
 or, after Fourier transformation to momentum space

\bea
 T_0 &=& \left[\frac{ 16\Phi^4}{(1-\Phi^2)^2}\right]^{N/4} \;
  e^{ \ds (2a+i\pi) \,\hat h_0 }, \NL
 T_1 &=& e^{ \ds -K \, \hat h_1 }, \NL
 T_{2,3} &=& e^{\ds -2\ell \,N/4} \;\; e^{\ds \frac{1}{2} \ell \, \hat h_{2,3}}
\eea
with
\bea
\hat h_0 &=&  \sum_{-\pi\leq k< \pi} \bh_k \bv_k\;, \label{hath0}\\
\hat h_1 &=&  \sum_{-\pi\leq k< \pi} \Big[ i \sin(k)\, \Big( \bh_k \bh_{-k}
  +  \bv_k\bv_{-k} \Big) 
  + \cos(k)\, \Big( \bh_k\bv_k -  \bv_k\bh_k \Big) \Big] \label{hath1}\\
\hat h_{2,3} &=&  \sum_{-\pi\leq k< \pi} 
  \Big[ i \sin(k)\,\Big( \bh_k \bh_{-k} + \bv_k \bv_{-k} 
\mp \bh_k\bv_{k+\pi} \mp  \bv_k\bh_{k+\pi} \Big)\NL   &{}& +  \cos(k)\,
\Big(\bh_k\bv_k -  \bv_k\bh_k \pm \bv_k \bv_{\pi-k} \mp \bh_k \bh_{\pi-k} \Big)\Big].
\label{hath23}
\eea

We work with periodic boundary conditions. For states with even (odd) numbers of fermions 
this requires that $b_N^{\dagger} =\bh_0$ ($\bh_N =-\bh_0$) (cf. reference \cite{SML64}). To
satisfy these conditions one chooses in $\bv_n = \frac{1}{\sqrt{N}} \sum_k e^{ikn} \; \bv_k$ 
 wave numbers $ k = \frac{2 \pi}{N} m$ or $k = \frac{2 \pi}{N} (m+1/2)$, $m$ integer,  
 for the even and odd states, respectively.\\   
The non-Bravais structure of the auxiliary lattice in figure \ref{aniskago} manifests itself 
in the structure of the hamiltonians $\hat{h}_{2,3}$: they couple not only fermionic states 
with wave numbers $k$ and $-k$ as in the square lattice case, but also states with wave 
numbers $k$ and $\pm k+\pi$. To deal 
with this situation we divide the Brillouin zone
 $-\pi \leq k < \pi$ 
into 4 subintervalls, $-\pi \leq k  < - \pi/2,\;\; -\pi/2 \leq k < 0;\;\;0 \leq k < \pi/2,\;\; 
\pi/2 \leq k < \pi$ and choose $N$ to be a multiple of $4$. Then,  $\hat h_0$, $\hat h_1$ and 
$\hat h_{2,3}$ are diagonal in $q$ and can be represented in the 
reduced Brillouin zone $0 \leq q < \pi/2$. For instance,

\bea
\hat h_0 &=& \ \sum_{0 < q < \pi/2} \hat h_{0q} + \hat h_{0b}^{even} \qquad \mbox{with} \NL
\hat h_{0q} &=&\bh_{-\pi + q} \bv_{-\pi + q} +
 \bh_{-q} \bv_{-q} +
 \bh_{q} \bv_{q} + \bh_{\pi - q} \bv_{\pi -  q} \;,\; \mbox{and} \NL
\hat h_{0b}^{even} &=& \bh_{-\pi}  \bv_{-\pi} + \bh_{-\pi/2} \bv_{-\pi/2} +\bh_0 \bv_0
  + \bh_{\pi/2} \bv_{\pi/2}\;.
 \label{hred}
\eea

The extra term $\hat h_{0b}^{even}$ corresponds to the values
 $k = -\pi,\; -\pi/2,\;
0, \pi/2$ of the original Brillouin zone. It occurs only if $\hat h_0$ acts in the space of 
 {\it even states}. Then,

\be
T_0^{even} = \left(\frac{16\Phi^4}{(1-\Phi^2)^2}\right)^{N/4} T_{0b} \prod_{0 < q < \pi/2} T_{0q},
\ee

where $T_{0b} = e^{2a\hat h_{0b}^{even}}$ and $T_{0q} = e^{2a\hat h_{0q}}$. Similarly, $T_1$ and 
$T_{2,3}$ are products of matrices acting in the subspaces of the individual wave numbers $q$. 
We remark that while 
these products consist of $N/4 - 1$ factors corresponding to the wavenumbers $0 < q < \pi/2$, 
there is only $1$ factor of the type $T_{0b}$ in each of the matrices $T_0$, $T_1$, 
$T_2$, 
and $T_3$. In the present paper we are always interested in the thermodynamic limit 
$N \rightarrow \infty$ of our results. In this limit the factors $T_{0b}$ play no role 
and will therefore be neglected in the sequel.

\section{Eigenvalues and eigenvectors of the symmetric transfer matrix}

Our next aim is to determine the largest eigenvalue and the corresponding eigenfunction 
of the symmetric transfer matrix (\ref{stf}). It consists of factors that are all exponentials 
of quadratic forms of fermion operators. Hence $T$ is also an exponential of a quadratic form of 
fermion operators. To determine this last exponential, the multiplication of the individual 
factors in  equation \ref{stf} has to be performed explicitly. To achieve this, 
we find it convenient to work 
with the representation of the fermion operators $b_q$, $b_q^{\dagger}$ by Grassmann variables 
$\gamma_q$, $\gamma_q^*$. In appendix A  we briefly describe the steps that are 
necessary to convert operators such as  $T_0,1,2,3$ into the so called 
{\it matrix form}. 
For details, we refer the reader to the book by F. A. Berezin \cite{Berezin66}, in particular to 
chapter 6. We introduce the shorthand notation 
 $\gvv_q = \{\gv_q,\gv_{-q},\gv_{-\pi + q},\gv_{\pi-q} \} \equiv 
\{\gv_1,\gv_2,\gv_3,\gv_4 \}$. (There is no need to distinguish the row vector $\gvv_q$ from 
its adjoint column vector $\gvv_q{}^t$  here, since dyadic products $\gvv_q \cdot \gvv_q{}^t$ will 
not occur.) In matrix form, $T_0$, $T_1$ and $T_{2,3}$ read (here and in  the sequel  we use 
the shorthand notation $\cos(q) = c_q$, $\sin(q) = s_q$ )

\be
\Ttnull(\ge,\gv) =  \left( \frac{4\Phi^2}{1-\Phi^2} \right)^2 \; 
 e^{\ds - h_0\; \gve_q \gvv_q} \;, \quad h_0 = \frac{1}{\Phi},  \label{tT0}
\ee

\vspace{2mm}
\be 
\Tteins(\ge,\gv) = \frac{(1+\Psi^2)^2-4c_q^2\Psi^2}{(1-\Psi^2)^2} \;\; \;
 e^{\ds \gve_q \, h_1\,\gvv_q \,+\,\frac{1}{2} \gve_q \,h_1'\, \gve_q
  \,-\,\frac{1}{2} \gvv_q \,h_1'^* \, \gvv_q} \label{tT1} 	
\ee
with
\be
 h_1 = \mm{cc}{\frac{1-\Psi^2}{1+\Psi^2+2c_q \Psi}&0\\
  0 & \frac{1-\Psi^2}{1+\Psi^2-2c_q \Psi}} , \quad
 h_1' = \frac{2 s_q \Psi}{1 - \Psi^2}\; h_1 %\mm{cc}{\frac{2s_q \Psi}{1+\Psi^2+2c_q \Psi}&0\\
  %0&\frac{2s_q \Psi}{1+\Psi^2-2c_q \Psi} }
  \mm{cc}{\sy&0\\0&-\sy} , \label{th1} 
\ee

\be
\Ttzwei(\ge,\gv) =  \frac{1}{4}\; 
 e^{\ds \gve_q\,(1+h_{2})\,\gvv_q \,+\,\frac{1}{2}\gve_q\,h_{2}'\,\gve_q
 \,-\,\frac{1}{2} \gvv_q \,h_{2}'^* \, \gvv_q} \label{tT23}
\ee

with

\be
 h_2 = \left( \begin{array}{cc}c_q&i s_q \sz\\ -i s_q \sz&-c_q \end{array} 
 \right) , \label{th2}\;\;
 h_2' = - \left(\begin{array}{cc}s_q&i c_q\sz\\i c_q\sz&s_q \end{array}\right)
 \left( \begin{array}{cc}\sy&0\\0&-\sy \end{array}  \right).\label{th2p}
\ee

Here the limit $\ell \to \infty$ has already been taken.\\
From the remarks after equation (\ref{stf}) of section 2 it follows that 

\bea
T &=& T_{1/2} \cdot T_{1/2},\quad \mbox{where} \NL
T_{1/2} &=& T_2\;T_0\;T_1\;T_0\;s\;T_2.  \label{Thalf} 
\eea

Thus it suffices to determine the eigenvalues and the eigenstates of $T_{1/2}$. 
Following the technique described in appendix A, we determine the {\it matrix form}
of $T_{1/2}$:
 
\bea
\widetilde{T_{1/2\;q}}(\ge\gv) &=& \int d\lambda^* d\lambda \; d \mu^* d \mu \; 
d \nu^* d \nu \; d \rho^* d \rho \;\; e^{\displaystyle -\lve \lvv} \; e^{\displaystyle -\mve \mvv} 
 \; e^{\displaystyle -\nve \nvv} \; e^{\displaystyle -\rve \rvv} \NL
 && \tilde T_2(\ge,\lambda) \;
 \tilde T_0(\lambda^*,\mu) \; \tilde T_1(\mu^*,\nu)\; \tilde T_0(\nu^*,\rho)\;
\tilde T_2(s \rho^*, \gv) \NL
&=&  v\;
e^{\displaystyle \gve_q \, H\,\gvv_q \,+\,
 \frac{1}{2} \gve_q \, H'\, \gve_q
  \,-\,\frac{1}{2} \gvv_q \, H'^* \, \gvv_q} \label{mfT}\\
\widetilde{T_{1/2}} &=& \prod_q \widetilde{T_{1/2\;q}}\;.
\eea

(We have suppressed the momentum label $q$ in the first two lines of this equation.)
The shift operator acts on $\rho^*$ as 
$s\;\rho^* = \{\rho_1^*,\;\rho_2^*,\;-\rho_3^*,\;-\rho_4^*\}$.\\
We point out here that being quadratic in fermion variables, the exponent in 
$\widetilde{T_{1/2\;q}}$, equation (\ref{mfT}), is the hamiltonian of a system of noninteracting
 fermions.  

The general structure of the hamiltonians $H$ and $H'$ derives from the structure of the 
hamiltonians $h_0$, $h_1$, $h'_1$, $h_2$ and  $h'_2$, equations (\ref{tT0}), (\ref{th1}), 
(\ref{th2}):  $h_0$, $h_1$, $h_2$ and $h'_1$, $h'_2$
 are of the form

\be 
\sM = \left(\begin{array}{cc} a & ib\,\sz\\ic\,\sz& d\end{array}\right)
\;\mbox{and}\;\;\;
\sM \cdot u,\;\;
 u = \left(\begin{array}{cc} \sy & \\ & -\sy\end{array} \right),\mbox{respectively}, 
\ee
where $a$, $b$, $c$, and $d$ are real scalars.   
Products of $\sM$ are again of the 
form $\sM$ and furthermore $[\sM,u] = 0$, $u^2 = 1$. Hence, in  equation (\ref{mfT}), $H$ is 
of the form $\sM$ and $H'$ is of the form $\sM\;u$.
Furthermore, since $\widetilde{T_{1/2}}$ is a real symmetric quadratic form in the space of the 
Grassmann vectors $\gvv_q$, $H$ must 
be symmetric and $H'$ must be antisymmetric. Hence $H$ and $H'$ must have the following structure: 
\be
H = \left(\begin{array}{cc} a & ib\,\sz\\ -ib\,\sz& d\end{array}\right)\;\; 
\mbox{and}\;\; H' = \left(\begin{array}{cc} a' & ib'\,\sz\\ib'\,\sz& d\end{array}\right)\;
 \left(\begin{array}{cc} \sy & \\ & -\sy\end{array} \right).
\label{HH'}
\ee
The real scalars  $a$,  $b$, $d$ and $a'$,  $b'$, $d'$ are rational functions of $\Psi$
and $\Phi$. Explicit expressions for these quantities are given in appendix B. Their 
calculation is straight-forward.\\ 

Turning to the eigenvalue problem for the transfer matrix we first consider the space of 
even states. As an even eigenstate we choose in Grassmann representation 

\bea
 \phi_g(\ge)&=& \phi_0 -i \phi_1 \;\ge_1 \ge_2 + i \phi_2 \;\ge_3 \ge_4
            - \phi_3 \; \frac{1}{\sqrt{2}} (\ge_1 \ge_4 + \ge_2 \ge_3)\NL
	    & & + \phi_4 \; \ge_1  \ge_2 \ge_3 \ge_4\;.
\label{Gphig}
\eea
  
Again the momentum variable $q$ is suppressed here. 
%We are of course free to set $\Phi_0 = 1$ and we shall do so lateron. 
Only  products of Grassmann variables with total momentum  
$q = 0\; \mbox{mod}\; \pi$ are included 
in this state. The coefficients  $\phi_{0,1,2,3,4}$ of the eigenstate (\ref{Gphig}) 
are calculated from

\be
\int \!\!d\le d\lv \;\; \widetilde{T_{1/2}}(\ge,\lv) \;\; 
 e^{-\le\lv} \;\; \phi_g(\le) = \sqrt{E_q^{(g)}} \; \phi_g(\ge),
\label{ewgeq}
\ee
where (see (\ref{HH'})) 

\bea
 \widetilde{T_{1/2}}(\ge,\lv) = v \exp\big\{&& 
 a (\ge_1 \lv_1 + \ge_2 \lv_2) + d (\ge_3 \lv_3 + \ge_4 \lv_4)\NL 
&&+ ib ( \ge_1 \lv_3 - \ge_2 \lv_4 - \ge_3 \lv_1 + \ge_4 \lv_2 )\NL
&&  
 -ia' \ge_1 \ge_2 + id' \ge_3 \ge_4  - b' ( \ge_1 \ge_4 + \ge_2 \ge_3 ) \NL 
&& -ia' \lv_1 \lv_2 + id' \lv_3 \lv_4  + b' ( \lv_1 \lv_4 + \lv_2 \lv_3 ) 
 \big\},
\label{Tgabd}
\eea
  
and we observe that the products with nonzero (mod $\pi$) momenta, $\ge_1 \ge_3$ and 
$\ge_2 \ge_4$, are indeed not generated by the action of $\widetilde{T_{1/2}}$ on $ \phi_g(\ge)$.

Equation (\ref{ewgeq}) results in

\be
 \mathcal{T}  \mm{c}{\phi_0\\\phi_1\\\phi_2\\\phi_3\\\phi_4 }
= \sqrt{E_q^{(g)}} \mm{c}{\phi_0\\\phi_1\\\phi_2\\\phi_3\\\phi_4}. 
\label{eigeq}
\ee

The $ 5 \times 5$ matrix $\mathcal{T}$ is presented in appendix B. Using the explicit
expressions for the elements $a$, $b$, $d$, $a'$, $b'$, $d'$ (see appendix B)  we find 

\be
 \det(\mathcal{T} -z) = -z^5 - 2 c_5 \; z^4 - c_4 \; z^3,
\label{detTz}
\ee

with

\bea
c_5 &=& -8\, c_q^2 \; \left[
 -\left( \frac{1-\Psi}{1+\Psi} \right)^2 
 \frac{1+6\Phi^2+\Phi^4}{(1-\Phi^2)^2} 
 + 2 \left( \frac{1+\Phi^2}{1-\Phi^2} \right)^2
 - \left( \frac{1+\Psi}{1-\Psi} \right)^2 \right] \NL 
&&\quad-8 \left[\frac{1-\Psi}{1+\Psi}\left( \frac{1+\Phi^2}{1-\Phi^2} \right)^2 
+ \frac{1+\Psi}{1-\Psi} \right]^2,\label{c5}\\
c_4 &=& 2^{14} \; c_q^2 \; \left( \frac{1-\Psi}{1+\Psi} \right)^2 
 \frac{(1+\Phi^2)^2 \;\Phi^4 }{(1-\Phi^2)^6}. \label{c4}
\eea

$c_5$ is negativ and hence the largest eigenvalue in (\ref{eigeq}) is
\be
\sqrt{E_q^{(g)}} = - c_5 + \sqrt{ c_5^2 - c_4}.
\label{eig}
\ee
\vspace{3mm}

As we need to know the largest eigenvalue of the transfer matrix  
for all positive values of $J/J'$, we have to also consider the space of odd states. 
The appropriate ansatz for an odd eigenstate is 

\bea
 \phi_u(\ge) &=& \bar{\phi}_1 \,\ge_1 + \bar{\phi}_2 \,i\ge_3 
 + \bar{\phi}_3 \,i\ge_1  \ge_3 \ge_4 + \bar{\phi}_4 \, \ge_1 \ge_2 \ge_3 \NL
 && \quad + \bar{\phi}_{-1} \,\ge_2 + \bar{\phi}_{-2} \,i\ge_4 
  + \bar{\phi}_{-3} \,i\ge_2  \ge_3 \ge_4 + \bar{\phi}_{-4} \,\ge_1 \ge_2 \ge_4\;. 
\label{ewu}
\eea

Here the products of Grassmann variables in the first four terms have total momentum 
$q\; \mbox{mod}\; \pi$, while the products in the last four terms have total momentum 
$-q\; \mbox{mod}\; \pi$. As the momentum $q$ is conserved (mod $\pi$)
in our model, the eigenvalue equation resulting from

\be
\int \!\!d\le d\lv \;\; \widetilde{T_{1/2}}(\ge,\lv) \;\; 
 e^{-\le\lv} \;\; \phi_u(\le) = \sqrt{E_q^{(u)}} \; \phi_u(\ge)
\label{ewueq}
\ee 
  
is block diagonal,

\be
 \mathcal{T}_{\pm}
\left(\begin{array}{c}\bar{\phi}_{\pm 1}\\ \bar{\phi}_{\pm 2}\\ \bar{\phi}_{\pm 3}\\ \bar{\phi}_{\pm 4} \end{array}\right)
= \sqrt{E_q^{(u \pm)}}
\left(\begin{array}{c}\bar{\phi}_{\pm 1}\\ \bar{\phi}_{\pm 2}\\ \bar{\phi}_{\pm 3}\\ \bar{\phi}_{\pm 4} \end{array}\right).
\label{eiueq}
\ee

Explicit expressions for the matrices $\mathcal{T}_+$, $\mathcal{T}_-$ are 
to be found in appendix B. The two largest  eigenvalues $\sqrt{E_q^{(u+)}}$, 
$\sqrt{E_q^{(u-)}}$ turn out to be identical,

\be
\sqrt{E_q^{(u+)}} = \sqrt{E_q^{(u-)}} = -c_4
\label{eiu}.
\ee

\vspace{3mm}

As a first application of the results obtained so far in this section, we identify the transition 
line between ordered and disordered states of the AIK. As is shown in section VB of 
reference \cite{SML64} a transition from  a high-temperature disordered phase to an ordered state 
at low temperatures occurs when the even and the odd eigenstates become degenerate as the 
temperature is decreased. Thus we have to compare the eigenvalues 
$E_q^g = (- c_5 + \sqrt{c_5^2 -c_4}\;)^2$ and $E_q^u = c_4$ as functions of the temperature $T$ 
in the different regions of $J/J'$.\\ 
{\it 1}) $J/J'<1$: Using the explicit expressions for $c_4$ and $c_5$, equations (\ref{c4}), 
(\ref{c5}), 
one finds after some algebra that $E_q^g > E_q^u$ for all $q \neq 0$. For $q=0$ $E_q^g$ and 
$ E_q^u$ become equal if

\be
\frac{1-\psi}{1+\psi} = \frac{1-\phi^2}{2 \phi}\quad \mbox{or} \quad 
  J = \frac{T_c}{2} \ln \frac{\sinh^2(\frac{2 J'}{T_c})}{2\cosh(\frac{2 J'}{T_c})}.
\label{trline}
\ee 

This defines the transition line between the ferrimagnetic and paramagnetic phase of the AIK 
(see figure \ref{phase-d} below) which was already determined by Kano and Naya.\\

{\it 2}) $J=J'$, isotropic kagome model: Here $E^{(g)}_q$ is larger than $E^{(u)}_q$ 
for all momenta $q$ and for all temperatures including $T=0$. This reflects the fact that 
the ground state of the Ising model on the kagome lattice is disordered.\\

{\it 3}) $J/J'>1$. Using the explicit expressions for $c_4$ and $c_5$ again, one finds that 
$E^{(u)}_q/E^{(g)}_q$ attains its maximum at $q=0$. As $T$ decreases to zero,
 $E^{(u)}_q/E^{(g)}_q \rightarrow 0$ for $q \neq 0$. However, for $q=0$, the ratio 
$E^{(u)}_q/E^{(g)}_q \rightarrow 1$ as $T \rightarrow 0$, implying that there is long 
range order in the ground state of the AIK.\\
These facts will be discussed in more detail below after we have determined the correlation 
function between the middle spins. \\
Since $E^{(g)}_q > E^{(u)}_q$ for all $q > 0$, the free energy per site of the AIK is

\be
f = -\frac{4}{N}\; T\!\!\!\!\sum_{0 \leq q < \pi/2} \ln(E^{(g)}_q)\;.
\label{freeerg}
\ee 

\begin{figure}[b]
\caption{\label{entr} Entropy $\mathcal{S}$ per site of the AIK model. $J'$ 
has been set to unity here. Temperatures (from bottom to top): $0.05$, $0.3$,  $0.5$,  $0.75$,  $1.0$,  $1.5$, $2.5$, 
$10.0$ in units of $J'$. As $T \to \infty $, the entropy $\mathcal{S}$ reaches its limit $\ln 2$. As $T \to 0$,
 $\mathcal{S}$ decreases to $0$ for $J < J'$ while a rest entropy of 
$\frac{1}{3} \ln2$ remains for $J > J'$ (see introduction).}
%\psfrag{ln 2}[][][2.5]{$\ln 2$}
%\psfrag{1/3 ln 2}[][][2.5]{$1/3 \ln 2$} 
%\psfrag{0.50183}[][][2.5]{$0.5018$} 
%\psfrag{S}[][][2.5]{$\mathcal{S}$}
%\psfrag{J}[][][2.5]{$J$}\psfrag{0.5}[][][2.5]{$0.5$}
%\psfrag{0}[][][2.5]{$0$}  
%\psfrag{1}[][][2.5]{$1$} \psfrag{1.5}[][][2.5]{$1.5$}
%\psfrag{2}[][][2.5]{$2$}
%\includegraphics[angle=-90,width=.5\textwidth]{entropie.eps}
\includegraphics[width=.5\textwidth]{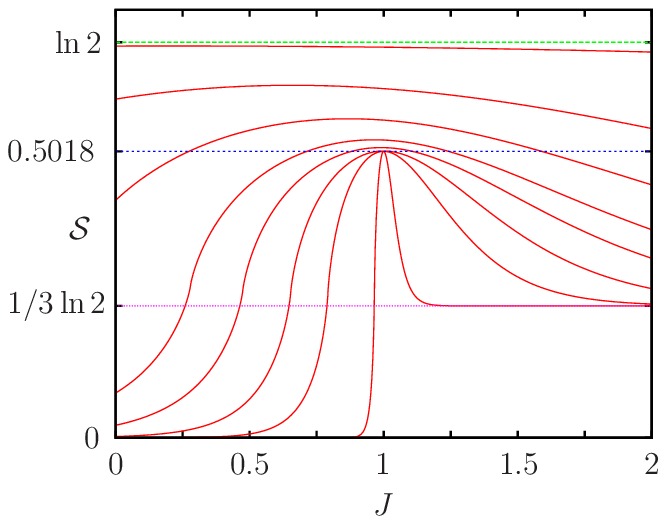}
\end{figure}

In figure \ref{entr} we show the entropy per site that results from this expression.

Obviously, the isotropic point $J=1$ plays an exceptional role in the thermodynamics of the AIK.

\section{Correlation function between middle spins}

For a system with $M$ rows of spins the correlation function between two middle spins at 
sites $m$ and $n$ ($m < n$) of the same row is given by

\be
\chi(m,n) = \frac{\tr{\, T_{1/2}^{\;2 M} \;\sz_m \sz_n\,}}
 {\tr{\, T_{1/2}^{\;2 M}\,}}
\label{cfsigM}
\ee

where $T_{1/2}$ is the transfer matrix (\ref{Thalf}) defined in section 3. If the 
state $|\suz\rangle$ with maximum eigenvalue of the transfer matrix  $T_{1/2}$ is non-degenerate 
this reduces to 

\be
\chi(m,n) = \frac{\langle \suz| \;\sz_m \sz_n\,|\suz\rangle} 
 {\langle \suz|\suz\rangle}
\label{cfsig}
\ee
 
in the thermodynamic limit $M \rightarrow \infty$. Again, the further development in this 
section follows closely the method devised by Schultz, Mattis and Lieb \cite{SML64} in 
section IV of their work.
Applying the Jordan-Wigner transformation, equation (\ref{szn}) to the product $\sigma^z_m\; \sigma^z_n$ 
we have 

\be
\sigma^z_m\; \sigma^z_n = i b_m^y b_n^x e^{i\pi \sum_{j=m+1}^{n-1} b_j^\dagger\;b_j},
\label{ss}
\ee

where

\be
b_j^x = b_j^\dagger + b_j \quad \mbox{and} \quad i b^y_j =  b_j^\dagger - b_j\;.
\label{cxcy}
\ee

Expanding the exponential on the r.~h.~s.~of equation (\ref{ss}) we obtain for the correlation 
function (\ref{cfsig})

\be
\chi(m,n) = \langle{\phi_g}| (i b_m^y)\,(b_{m+1}^x)\;(i b_{m+1}^y) \cdots (b_{n-1}^x)\;
(i b_{n-1}^y)\;(b_{n}^x)|\phi_g\rangle /\langle \phi_g |\phi_g \rangle .
\label{fchi}
\ee  

Here, $|\phi_g\rangle$ is the ground state of $-\ln T_{1/2}$ in the even 
subspace in the fermionic representation (c.~f.~equation (\ref{Gphig}),

\bea
|\phi_g\rangle &=& \prod_{q=0}^{\pi/2}\;|\phi_g\rangle_q, \NL
\mbox{where}\NL
|\phi_g\rangle_q &=& \phi_0(q) -i \phi_1(q) \; b_{1q}^{\dagger} b_{2q}^{\dagger}  
+ i \phi_2(q) \;b_{3q}^{\dagger} b_{4q}^{\dagger}\NL  
& &-\phi_3(q) \; \frac{1}{\sqrt{2}} ( b_{1q}^{\dagger} b_{4q}^{\dagger}  + 
b_{2q}^{\dagger} b_{3q}^{\dagger})
+\phi_4(q)\; b_{1q}^{\dagger} b_{2q}^{\dagger} b_{3q}^{\dagger} b_{4q}^{\dagger}|0\rangle
\label{fevg}\\
(|0\rangle: & & \mbox{fermion vacuum}). \nonumber
\eea
   
The fermion operators $b_{1q}^{\dagger}$, $b_{2q}^{\dagger}$, $b_{3q}^{\dagger}$ and 
$ b_{4q}^{\dagger}$ act in the intervals $[0,\pi/2[$\;, $ [-\pi/2,0[$\;, 
$[-\pi , -\pi/2[$\; and $[\pi/2 \pi[$\;, respectively. Our choice of the even 
ground state in equation (\ref{fchi}) implies that by evaluating this expression we will obtain 
the correlation function in the disordered region of the phase diagram where $|\phi_g\rangle$
is the ground state. Indeed, we expect to find the most interesting results in this region.\\

Since $|\phi_g\rangle$ is the ground state of a system of noninteracting fermions, cf.~section 3, 
the expectation value in equation (\ref{fchi}) can be evaluated by 
the use of Wick's theorem, which requires the knowledge of the contractions of all pairs of 
operators of the product in equation (\ref{fchi}. In appendix B the coefficients $\phi_\nu(q)$, 
$\nu = 0,\cdots, 4$ are found to be real for all $q$. This suffices to show that 
\be   
\langle \phi_g|b^x_j\;b^x_l| \phi_g \rangle = 0 \quad \mbox{and} \quad
\langle \phi_g|b^y_j\;b^y_l| \phi_g \rangle = 0 
\label{conxxyy}
\ee
for all pairs of sites $j \neq l$. Denoting the remaining contractions by $G_{j,l+1}$,

\be
G_{j,l+1}=\langle \phi_g|i\;b^y_j\;b^x_{l+1}| \phi_g \rangle/\langle \phi_g|\phi_g \rangle,
\label{conxy}
\ee

we can thus cast the result for the correlation function (\ref{fchi}) into the form 

\be
\chi(m,n) = \left|\begin{array}{lllll} G_{m,m+1} & G_{m,m+2} & \cdots &  G_{m,n-1} &  G_{m,n}\\
 G_{m+1,m+1} &  G_{m+1,m+2} & \cdots & G_{m+1,n-1} &  G_{m+1,n}\\
\cdot & {} & {} & {} & \cdot \\
\cdot & {} & {} & {} & \cdot \\
\cdot & {} & {} & {} & \cdot \\
\cdot & {} & {} & {} & \cdot \\
\cdot & {} & {} & {} & \cdot \\
\cdot & {} & {} & {} & \cdot \\
G_{n-1,m+1} & {} & \cdots & G_{n-2,n} & G_{n-1,n}
\end{array} \right|.
\label{detchi}
\ee

In appendix C we find that $G_{j,l} = 0$, if  $j$ and $l$ are both even, or both odd, and that
$G_{j,l} = \delta_{j+1,l}$, if $j$ is odd. Furthermore, the remaining elements $G_{j,l}$ 
are found to depend only on the difference $l-j$. Hence, denoting   

\be
G_{2j,2l+1} = a_{l-j}
\label{defa}
\ee

we get for the correlation function between middle spins at the sites $1$ and $2n+1$
\bea
\chi(1,2n+1) &=&\left|\begin{array}{llllcll} 1\qquad & 0\qquad & 0\qquad & 0\qquad & \cdots & 0\qquad  &  0\\
 0 &  a_0 & 0 & a_1 & \cdots & 0 &  a_{n-1}\\
 0 & 0 & 1 & 0 & \cdots & 0 & 0 \\
 0 & a_{-1} & 0 & a_0 & \cdots & 0 & a_{n-2} \\
 0 & 0 & 0 & 1 & \cdots & 0 & 0 \\
 \cdot & {} & {} & {} & {} & {} & \cdot \\
 \cdot & {} & {} & {} & {} & {} & \cdot \\
 0 & 0 & 0 & 0 &  \cdots & 1 & 0 \\
 0 & a_{-n+1} & 0 & a_{-n+2} &\cdots & 0 & a_0
\end{array} \right| \nonumber\\[3mm]  
& = &\left|\begin{array}{llcl} a_0\quad &a_1\quad& \cdots & a_{n-1}\\
 a_{-1} & a_{0} & \cdots & a_{n-2}\\
 \cdot & {} & {} \cdots & {}\\
  \cdot & {} & {} \cdots & {}\\
  \cdot & {} & {} \cdots & {}\\
 a_{-n+1} &  a_{-n+2} & \cdots & a_0
\end{array}
\right|\;.
\label{toepchi}
\eea

Thus, for the AIK, the correlation function between the middle spins is represented by a 
Toeplitz determinant, just like the spin-spin correlation function of the square lattice 
Ising model, cf.~reference \cite{SML64}.\\
In appendix C we show that   

\be
 a_j =  \oint \limits_{|z|=1} \!\!\frac{dz}{2\pi i} \;\frac{1}{z} \;
 \frac{1}{z^{j}} \;\cdot\; \frac{ p_2 + p_1 z^{-1} + p_0 z^{-2} }
{|p_2 + p_1 z^{-1} + p_0 z^{-2}|} , 
\label{cotoep}
\ee

where

\bea
 p_0 &=& \left( 1-\Phi^2 \right)^4 \, \Psi^2, \NL
 p_1 &=&  - \left( 1-\Phi^2 \right)^2 \; \left[ (1-\Phi^2)^2 \; (1+\Psi^4) + 
 4 \Phi^2 \; (1-\Psi) \; (1-\Psi^3)\right],  \NL
 p_2 &=&  \left[(1-\Phi^2)^2 \; \Psi - 2 \Phi^2 \; (1-\Psi)^2  \right]^2.
\label{cointcotoep}
\eea

This expression for the elements of the Toeplitz determinant (\ref{toepchi}) agrees 
with the corresponding expression in reference \cite{G74} apart from a sign difference 
in $p_1$, which is probably due to a misprint in \cite{G74}.\\
As is described in detail in Chapter XI of reference \cite{McCoyWu}, for large $n$ the Toeplitz 
determinant (\ref{toepchi}) can be evaluated analytically by applying a technique that
was developed to solve Wiener-Hopf sum equations. We apply this technique here, but 
refrain from presenting  any details  apart from the remark that, as in the case of the 
square lattice Ising model, we have to deal with the case of {\it index $\nu=-1$} here 
(see reference \cite{McCoyWu}, p.~250). The result takes the form 

\be
\chi(1,2n+1) = (-)^n \; R(n+1) \; x_n^{(n)} 
\label{formchi}
\ee    

Asymptotically,

\be
(-)^n\,R(n)  \sim   - \left[ \frac{(1-z_B^2)(1-z_A^{-2})}{(1-z_B z_A^{-1})^2} \right]^{1/4}\\
\label{asymchi1}
\ee
and
\be
x_n^{(n)}  \sim  -\oint \limits_{|z|=1} \!\!\frac{dz}{2\pi i} \; z^{n-1} \;
 \sqrt{\frac{1-z\;z_B}{1-z\;z_A^{-1}}}
 \sqrt{\frac{1-z^{-1}\;z_B}{1-z^{-1}\;z_A^{-1}}}. 
\label{asymchi2}
\ee

(Here and in the sequel the relation symbol ``$\;\sim\;$'' implies that corrections to the
right hand side of the expressions vanish in the limit $n \rightarrow \infty$.) The real 
quantities
$z_A$ and $z_B$ are the roots of the the quadratic form $p_2 z^2 + p_1 z +p_0$ 
(see equation (\ref{cotoep})),

\be
z_{A,B} = -\frac{p_1}{2p_2} \pm \sqrt{\left( \frac{p_1}{2p_2} \right)^2 -\frac{p_0}{p_2}}\;.
\label{zAB}
\ee

Since $p_1 < 0$, the roots $z_A$ and $z_B$ are positive and  $z_B \leq 1 \leq z_A$ for all values 
of the parameters $\Phi$ and $\Psi$ outside the ordered region. However, depending on 
the values of 
$K=J/T$ and $K'=J'/T$
either \ding{172} $ z_B < z_A^{-1}$  or \ding{173} $z_B > z_A^{-1}$.\\

\vspace{5mm}
\noindent
\ding{172} $z_B<z_A^{-1}$\\[3mm]
We evaluate the integral in equation (\ref{asymchi2}) by deforming the integration 
contour such that it 
encircles  the branch cut of the integrand which extends from $z_B$ to $z_A^{-1}$,

\be
-x_n^{(n)} \sim \int_{z_B}^{z_A^{-1}} \!\frac{dt}{\pi} \; t^{n-1} \;
 \sqrt{\frac{1-t\;z_B}{1-t\;z_A^{-1}}\; \frac{t-z_B}{z_A^{-1}-t}}\;.
\label{asymx1}
\ee

In the limit of large $n$ we obtain (see equation (\ref{formchi}))

\be
\chi(1,2n+1) \sim \left[ \frac{(1-z_B^2)(1-z_A z_B)^2}{1-z_A^{-2}} \right]^{1/4} 
  \cdot \sqrt{\frac{1}{\pi n}} \; z_A^{-n} \;.
\label{finchiI}
\ee
\vspace{5mm}
\ding{173} $z_A^{-1}<z_B$\\[3mm]
 Proceeding  in the same manner as in case \ding{172} we  find
\be 
-x_n^{(n)}  \sim   \int^{z_B}_{z_A^{-1}} \!\frac{dt}{\pi} \; t^{n-1} \;
 \sqrt{\frac{1-t\;z_B}{1-t\;z_A^{-1}}\; \frac{z_B-t}{t-z_A^{-1}}}\;\NL
\ee
and hence in the limit of large $n$
\be
\chi(1,2n+1) \sim  - \left[\frac{(1-z_B^2)^3 (1-z_A^{-2})}{(1-z_B z_A^{-1})^4} 
 \left(\frac{z_A z_B}{z_A z_B -1} \right)^2 \right]^{1/4}\!\!\!\! \cdot \sqrt{ \frac{1}{4\pi n^3}} z_B^n\;.
\label{finchiII}
\ee

\vspace{5mm}
We remark here that for the expressions (\ref{finchiI}) and (\ref{finchiII}) to be valid, 
the distances $n$ must be large enough so that the condition

\be
n \gg \bar{n} \equiv \left| \ln(z_A z_B) \right|^{-1} 
\label{condzAzB}
\ee

is satisfied.

\begin{figure}
\includegraphics[width=.5\textwidth]{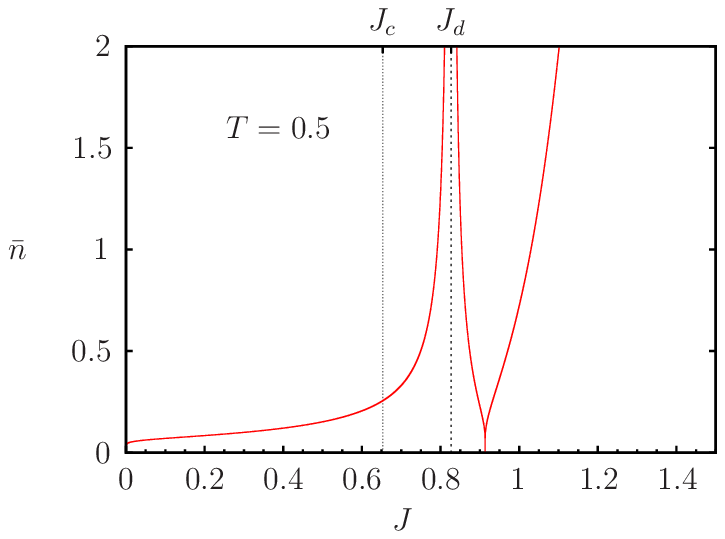}
\caption{\label{difflnzAB} $\bar{n}$ ($J' = 1$); as $J \to 0.91$, $\bar{n} \to 0$ since
 $z_A z_B \to 0$ in this limit, cf. equation (\ref{condzAzB}).}
\end{figure}
The plot of $\bar{n}$ in figure \ref{difflnzAB} indicates that in 
large regions of the phase diagram, in particular for $J > 1$, the correlation function
 $\chi(1,2n+1)$ attains its asymptotic behaviour  only for rather large $n$.

\section{Results and discussion}

\begin{figure}[b]
\caption{\label{phase-d} Phase diagram ($T$ in units of $J'$, $J'$ is set to unity). 
FM: ferrimagnetically ordered region, full line: critical line $J_c(T)$;\;\;\ding{172}, \ding{173} disordered regions, separated 
by the disorder line $J_d(T)$ (dashed line).}
\includegraphics[width=0.6\textwidth]{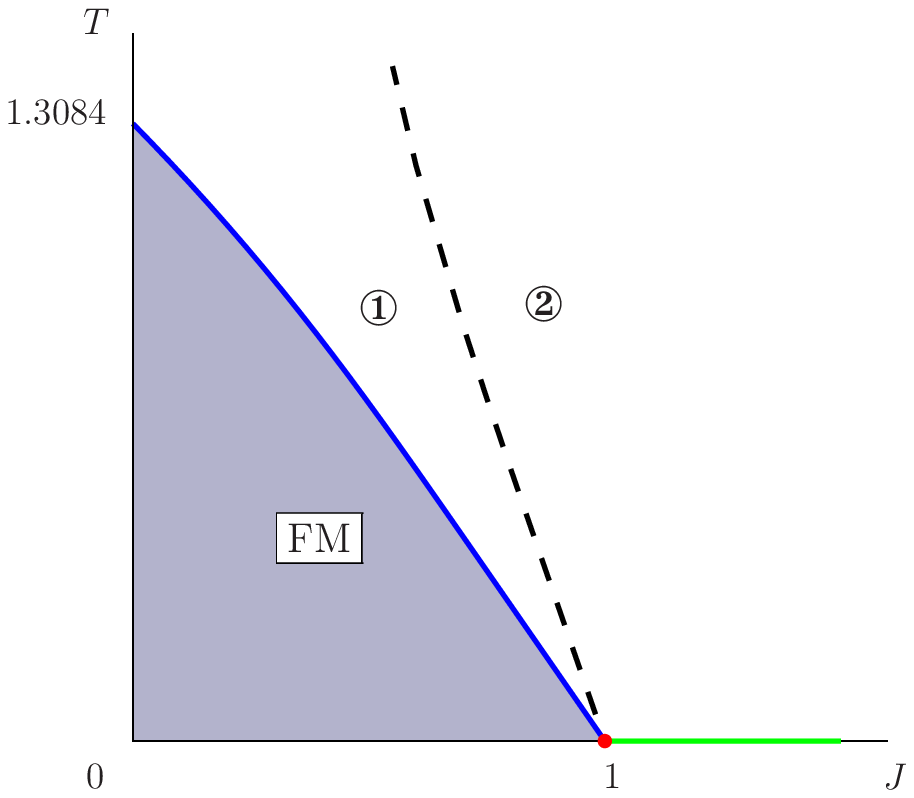}
\end{figure}

The most remarkable feature of the phase diagram, figure \ref{phase-d}, is the 
disorder line $J_d(T)$, 
which 
separates the regions \ding{172} and \ding{173} defined in the previous section from each 
other, see reference \cite{Rujan85} and references therein. It is determined  by the equation

\be
\Phi^2 = (\tanh(J'/T))^2 = \tanh(J/T) = \Psi\;.
\label{doline}
\ee

On this line 

\be
\frac{ p_2 + p_1 z^{-1} + p_0 z^{-2} }
{|p_2 + p_1 z^{-1} + p_0 z^{-2}|}  = \frac{1}{z}. \qquad (|z| = 1)\nonumber
\ee

Then it follows from \eq{cotoep} that in this case $a_j =  \delta_{j,-1}$ and hence 
$\chi(1,2n+1) = 0$, {\it i.e.} the middle spins are uncorrelated on the disorder line.

\subsection{Unfrustrated Ising like region}

In region \ding{172}, the asymptotic behaviour of the correlation function
is given by 

\be
\chi(1,2n+1) \sim C_1\!\cdot\!\frac{e^{-\frac{2n}{\xi_1}}}
{\sqrt{n}},\quad \xi_1^{-1}= \frac{\ln(z_A)}{2}, 
\;\; C_1 > 0\; (\mbox{cf.~equation (\ref{finchiI}})).
\label{chiising}
\ee

\begin{figure}
\includegraphics[width=.5\textwidth]{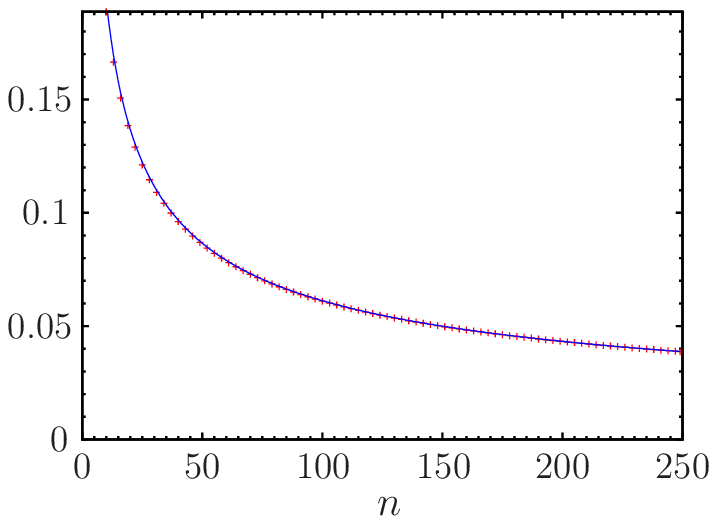}
\includegraphics[width=.5\textwidth]{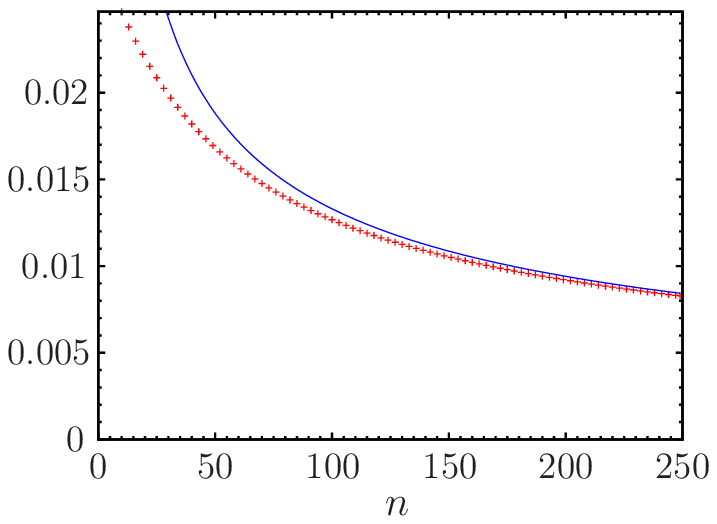}    
\caption{\label{preex7080} Comparison of the asymptotic (line) and the numerical results 
(symbols) for $\chi(1,2n+1) $ for $J=0.7, \bar{n}=0.3$ (left)  
and $J=0.825, \bar{n}=17.5$ (right) after division by $ e^{-2n /{\xi_1}} $, see text ($J'=1$). } 
\end{figure}

This suggests that in this region the AIK model behaves like a unfrustrated  
Ising magnet in the disordered phase, 
cf.~reference \cite{McCoyWu}, Chapter XI. In fact,
for $J=0$ the AIK model, cf.~figure \ref{aniskago}, reduces to the antiferromagnetic Ising model 
on the square lattice whose edges are decorated by the chain spins, which are decoupled from 
each other. Tracing out these spins yields an  effective ferromagnetic coupling  between 
the middle spins 
with coupling constant $K_{\Box}= - \frac{1}{2}\ln\cosh(2K')$.
Furthermore, for $J=0$
equations (\ref{zAB}) and (\ref{cointcotoep}) yield
  
\be
z_B = 0 \quad \mbox{and} \quad z_A^{-1} = -\frac{p_2}{p_1} =\sinh^2(2 K_{\Box}).
\ee
Then, the expression (\ref{finchiI}) for the correlation function 
between the middle spin of the AIK model  
reduces to the known expression for the correlation function between spins along one of 
the diagonals of the ensuing square model in the disordered region, cf.~reference \cite{McCoyWu}, 
Chapter XI.2.\\ 
To check the quality of the asymptotic expression (\ref{finchiI}) for the correlation 
function $\chi(1,2n+1)$ we have calculated the determinant in equation (\ref{toepchi}) numerically. In 
figure \ref{preex7080} the asymptotic result (\ref{finchiI}) and the numerical result are compared 
after the exponential factor $e^{-2n/\xi_1}$ has been divided out from both results. 
As is expected on account of the condition (\ref{condzAzB}), for small and moderate distances $2n$
the agreement 
is worse if $J$ is close to the disorder line where $\bar{n}\to\infty$.  
 
On the critical line $T_c(J)$, cf. equation (\ref{trline}), we find $z_A = 1$ and 

\be
 a_j = - \oint \limits_{|z|=1} \!\!\frac{dz}{2\pi i} \;\frac{1}{z} \;
 \frac{1}{z^{j}} \;\cdot\; \sqrt{\frac{-1}{z^2} \; \frac{z-z_B}{1-z\,z_B}}\;. 
\label{ajcrit} 
\ee

This is very similar to the corresponding expression for the square lattice Ising model,
equation (5.1) in Chapter XI of reference \cite{McCoyWu}. Applying the same method as McCoy and Wu,
 we arrive at

\be
\chi(2n+1,1) \sim  \left[ (1-z_B)^2 (1-z_B^2) \right]^{(1/4)} \frac{A}{n^{(1/4)}} 
\;\; (A: \mbox{see reference}\;\cite{McCoyWu}) 
\label{chicrit}
\ee

where 

\be
 z_B = \frac{(1+2\Phi^2-\Phi^4)^2 (1-2\Phi^2-\Phi^4)^2}
        {(1+8\Phi^2+10\Phi^4+8\Phi^6+\Phi^8)^2},\quad A=0.645\cdots\;\;.
\label{zBcrit}
\ee
In summary, the results presented here  indicate clearly  that in region \ding{172} the 
features of the AIK model differ only quantitatively from  those of an  unfrustrated Ising 
model with antiferromagnetic nearest neighbour interaction. The  middle spins of one row 
of the model are correlated ferromagnetically. The appearance of a 
ferrimagnetically  ordered region for small coupling $J$ and low temperatures fully agrees 
with this picture of an unfrustrated model.

\subsection{Chain dominated region}

The  features of the AIK model change qualitatively when the coupling $J$ is increased
across the disorderline into the region \ding{173}. The disordered phase extends down to
$T = 0$ in this region, and the correlation function of the middle spins changes from the 
expression in equation (\ref{chiising}) to

\bea
\chi(1,2n+1) &\sim&  - C_2
 \cdot  \frac{e^{-\frac{2n}{\xi_2}}}{\sqrt{n^3}}, \label{chichain} \\
\mbox{with} &{}&\xi_2^{-1} = -\frac{\ln(z_B)}{2},\quad C_2 > 0 \quad (\mbox{see equation 
(\ref{finchiII}})).\nonumber
\eea 

This implies that all the middle spins in one row of the AIK model are predominantly correlated 
antiparallel to one another. In Appendix D we present a simple perturbative evaluation of 
the correlation function which is valid for  $J' \ll J$ and $J'/T \ll 1$. In this limit, 
the correlations 
between the middle spins of one row are mediated by the spins of  
 two adjacent chains which are correlated antiferromagnetically and to which the middle spins 
are weakly coupled. This mechanism leads quite naturally to a negative correlation between the 
any pair of middle spins of the same row.\\
As we have mentioned in the Introduction, for $J > J'$ the middle spins decouple from the 
chain spins  in the limit $T \rightarrow 0$  and become therefore completely uncorrelated.
To see how this happens as the temperature is decreased to zero we expand $z_A$ and $z_B$ 
(cf.~equation (\ref{zAB})) for low temperatures, i.~e.~$\epsilon=2e^{-2J'/T} \ll 1$,\quad  
$\xi_2^{-1} = e^{-2(J-J')/T} \ll 1$, 
\bea
 z_A^{-1} &=& 1 - 2\xi_2^{-1} + \frac{3}{2} \xi_2^{-2} 
  + \mO(\xi_2^{-1}\epsilon, \xi_2^{-2} \epsilon, \xi_2^{-3}),\NL
 z_B &=& 1 - 2\xi_2^{-1} + \frac{5}{2} \xi_2^{-2} + 
\mO(\xi_2^{-1}\epsilon,\xi_2^{-2}\epsilon,\xi_2^{-3}).
\eea

\begin{figure}
\includegraphics[width=.5\textwidth]{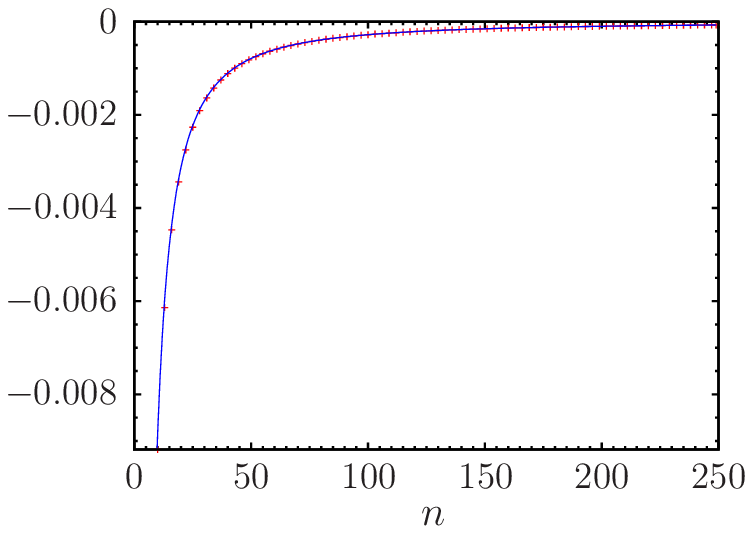}
\includegraphics[width=.5\textwidth]{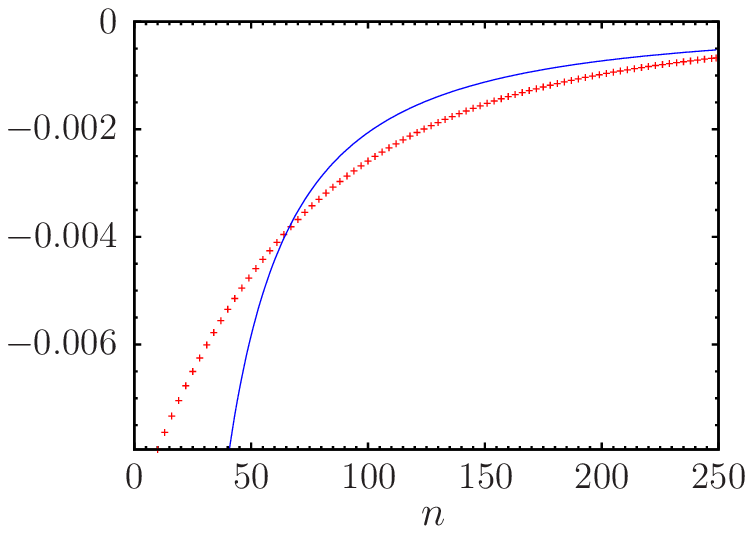}    
\caption{\label{preex91150} Comparison of the asymptotic (line) and the numerical results 
(symbols) for $\chi(1,2n+1)$ for $J=0.91, \bar{n} = 0.1$ (left) 
and $J=1.5, \bar{n}= 54.4$ (right) after division by $e^{-2n/\xi_2}$ ($J'=1$).} 
\end{figure}

With these expansions, equation (\ref{finchiII}) yields
\be
 \chi(1,2n+1) \sim  -\frac{1}{2}\;
 \xi_2^{-2} \; e^{ -2n/\xi_2}\;.
\label{chiIILT}
\ee
Although the correlation length $\xi_2$ tends to infinity as $T \rightarrow 0$, $ \chi(1,2n+1)$ 
vanishes in this limit since the prefactor $\xi_2^{-2}$ vanishes.\\
To compare the asymptotic result (\ref{finchiII}) for $\chi(1,2n+1)$ in the parameter region 
\ding{173} with the exact result (\ref{toepchi}) we have evaluated the latter 
numerically as in the other regime.
The comparison is shown in figure \ref{preex91150}.
As in region \ding{172}, the agreement between the asymptotic and the exact results becomes
 worse for small and moderate distances $n$ as $\bar{n}$ increases.

\subsection{The point $J' = J = 1$, $T = 0$}

From our discussion of the rest entropy of the AIK model in the introduction, it follows  
that in the limit $T \rightarrow 0$ this model behaves  irregularly at the isotropic 
point $J' = J = 1$. Here, we wish to discuss the behaviour of the correlation function 
between the middle spins as the isotropic point is approached from different directions 
in the phase diagram. We fix a direction by choosing a fixed value for the parameter $\alpha$ 
in the equation 

\be
 J - J' = \alpha T, 
\label{rays}
\ee

which represents straight lines with different slopes in the phase diagram that 
intersect at the isotropic point. For instance,  $\alpha = - \ln 2$ is the slope of 
the phase boundary $T = T_c(J-J')$ at $J' = J = 1$, $\alpha = -\frac{1}{2} \ln 2$ is 
the slope of the disorder line at this point and $\alpha = 0$ is the line perpendicular 
to the abscissa in $J' = J = 1$. By parametrising 

\be
 \Phi = 1 - \epsilon \qquad \Psi = 1 - \epsilon \; \sqrt{s} \quad \mbox{where}
\quad s = e^{-4\alpha}
\ee

we obtain from equations (\ref{cointcotoep}), (\ref{zAB})

\be 
 z_A^{-1} = z_B \; \left[ \frac{2-s}{2} \right]^2,\;\; 
 z_B = \frac{4+6s}{(2-s)^2} 
 - \sqrt{\left[\frac{4+6s}{(2-s)^2}\right]^2 - \left[ \frac{2}{2-s} \right]^2}\;\;.
\ee  

This yields for the different cases:

\begin{tabular}{llll}
$s= 16$:&  on the phase boundary & $\!\!\!\chi$ as in equation~(\ref{chicrit}), &  
$\!\!\!z_B = 1/49$\\     
$16 > s > 4$:& in  region \ding{172} & $\!\!\!\chi$ as in
equations (\ref{finchiI}),(\ref{chiising})& 
 $\!\!\!z_A^{-1} > z_B$\\  
$s = 4$:& on the disorder line &$\!\!\!\chi=0$ &$\!\!\! z_A^{-1} =  z_B = 1$\\
$4> s >0$: & in region \ding{173} & $\!\!\!\chi$ as in 
equations (\ref{finchiII}), (\ref{chichain})&
  $\!\! z_B > z_A^{-1}$\\
$s=1$:& on the perpendicular & $\!\!\!\chi$ as in equation (\ref{chichain})& 
 $\!\xi_2 \simeq 1.25$\\
$s = 0$:& on the line $T = 0$ & $\!\!\!\chi = 0$ cf.~equations (\ref{formchi}), 
& $\!\!z_A^{-1} =  z_B = 1$\\
{}& {} &\quad\quad\; (\ref{asymchi1}) and (\ref{asymchi2}) &$\!\!\!$ as on the \\
{}& {} & {} & $\!\!\!$ disorder line
\end{tabular}

\section{Summary and outlook}

In this paper we have adapted  the transfer matrix technique devised in 
reference \cite{SML64} for the solution of the  square lattice Ising model to 
the case of the Ising model on the anisotropic kagome model (AIK). The isotropic  kagome 
model is fully frustrated, but as the coupling $J$ in one of the three lattice directions is 
varied from zero to infinity, the model changes from the unfrustrated square lattice model 
to a set of antiferromagnetic chains. We have calculated the pair correlation function of 
middle spins, i.~e.~of spins on intermediate rows between the chains with coupling $J$,
(i) asymptotically exact, (ii) numerically and for the case $J \gg 1$ (iii) by a perturbative 
method. We find that the phase diagram of the AIK model is divided into two sections by a 
disorder line $J_d(T)$. On this line, the correlation between the middle spins vanishes exactly. 
The following statements need to be corroborated by a calculation of the correlations between 
the chain spins, but this was not a subject of the present paper. For 
$J < J_c(T)$ the model is ferrimagnetically ordered: we expect the chain spins and the middle 
spins to be  ordered ferromagnetically but antiparallel relative to each other;
in the region  $J_c(T) < J < J_d(T)$ we expect short range order of the same kind. (Short range 
ferromagnetic order between the middle spins is found in the present paper.) In summary, 
for $J < J_d(T)$ we expect the AIK model to behave like an unfrustrated Ising model.\\
At the disorder line, the correlation between the middle spins changes qualitatively. 
For $J_d(T) < J$ there is no long range order at any temperature. The three methods 
(i - iii) show that the pair correlation function of the middle spins is negative throughout this 
region. It decays exponentially with a correlation length $\xi_2(T)$ that diverges in the limit 
$T \to 0$. 
However, as $T \to 0$ the amplitude of the correlation function becomes 
proportional to $\xi_2(T)^{-2}$, i.~e.~ the middle spins become uncorrelated as $T \to 0$. Since 
the chains are antiferromagnetically ordered  at $T=0$, the loss of correlation between the 
middle spins was to be expected.\\
To complete the picture of the spin correlations of the AIK model various correlation functions 
in addition to the middle spin correlation function need to be calculated. We expect to be able 
to perform the necessary calculations by applying the transfer matrix technique devised here.

\section*{Acknowledgement}
\input{ack}

\appendix
\section{Working with Grassmann variables}

Here we collect rules to be followed in transforming 
operators from their fermion representation, e.~g.~expressions (\ref{tf0}) \dots (\ref{tf3}) 
for the transfer matrices $T_0$ \dots $T_3$, into their representation by Grassmann
variables and of the rules that were important in our manipulations of such operators.
We adhere to the terminology of reference \cite{Berezin66}.\\
If $\hat T(b^{\dagger}, b)$ is a normal ordered fermionic operator ( $b^{\dagger}$, $b$: fermion 
creation and annihilation operators), then 
$T(\ge, \gv) := \left. \hat T \right|_{b^{\dagger} \to \ge\!,\,b \to \gv}$,
where $\gamma$ are  Grassmann numbers and  $\gamma^*$ their complex conjugates, is called 
the ``normal form'' of $T$, and 
$\tilde T(\ge, \gv) := T(\ge, \gv) \; e^{\ge \gv}$\\
is called the ``matrix form'' of $T$. ($\bh$, $\bv$ and  $\ge$, $\gv$ can be vectors of 
fermion operators or Grassmann numbers, respectively.)\\
Gaussian integral: For an arbitrary $N \times N$ matrix $H$ we have\\[3mm] 

\bea
\prod_n \int d\ge_n d\gv_n \;\; 
e^{\ds -\ge_n H_{nm} \gv_m} = \det H. \\(\mbox{In the exponent, automatic summation over the 
indices is implied.})\nonumber
\label{gauss}
\eea   
Gaussian integral with external sources: for an antisymmetric $2N \times 2N$ matrix $R$:

{\arraycolsep1pt
\bea
\int d\le d\lv \; e^{\frac{1}{2}\mm{cc}{ {\sss\lv}&{\sss\le}} {\ds R^{-1}}
 \mm{c}{{\sss\lv} \\{ \sss\le}}
  + \mm{cc}{\sss\gv&\sss\ge} \mm{c}{\sss\lv \\ \sss\le} }\NL
\qquad\qquad\qquad\qquad\qquad\qquad = \pm \sqrt{\det R^{-1}}\;\; 
 e^{\frac{1}{2} \mm{cc}{\sss\gv&\sss\ge} {\ds R} \mm{c}{\sss\gv \\ \sss\ge} }
\label{gausssouce}
\eea}

Here, $\gv$ and $\lv$ are $N$-dimensional Grassmann vectors. The determinant of an 
antisymmetric matrix is always positiv. The sign of the square root depends on the order 
of the Grassmann integrations. It is irrelevant for our purposes. \\[3mm]
Trace: \be \tr{\hat T} = \int d\gv d\ge \; e^{\ge \gv} \; \tilde T(\ge, \gv) \label{spur} \ee. 
 
Product of fermionic operators: 
\be \hat A \cdot \hat B = \hat C: \quad 
\tilde C(\ge \gv) = \int d\le d\lv \; \tilde A(\ge \lv) \tilde B(\le \gv)
 \; e^{-\le \lv}.
\label{Gprod} \ee
In our case all products in question can be evaluated explicitly since the integrals 
that have to be performed are gaussian (cf.~equation (\ref{mfT})).

\section{The matrices $\tilde{\mathcal{T}}$, $\tilde{\mathcal{T}}_\pm$} 

As the matrices $\mathcal{T}$, $\mathcal{T}_+$ and $\mathcal{T}_-$ play a central 
role in our analysis of the AIK model, we list their elements explicitely here in terms 
of our basic variables $\Psi$ and $\Phi$ (cf.~equation (\ref{phipsi})). $\mathcal{T}$ 
is a real symmetric  
$5 \times 5$ matrix and $\mathcal{T}_{\pm}$ are real symmetric $4 \times 4$ matrices.
Defining  $\bar{\mathcal{T}}$ and $\bar{\mathcal{T}}_{\pm}$  by 

\be
\mathcal{T} = v\cdot\bar{\mathcal{T}},\quad \mathcal{T}_{pm} = v\cdot\bar{\mathcal{T}}_{pm}
%\quad \mbox{and} \mathcal{T}_- = v\cdot\bar{\mathcal{T}}_-
\label{barT}
\ee  

we have 
\def\deet{\left(a' d' + b'^2 \right)}
\bea
\bar{\mathcal{T}}_{11}=1,\quad \bar{\mathcal{T}}_{12}=a',\quad \bar{\mathcal{T}}_{13}=d',
\quad \bar{\mathcal{T}}_{14}=\sqrt{2} b',\quad \bar{\mathcal{T}}_{15}=a'd'+b'^2,\NL[2mm]
\bar{\mathcal{T}}_{22}=a^2+a'^2,\quad \bar{\mathcal{T}}_{23}=a' d' - b^2,\quad 
\bar{\mathcal{T}}_{24}=\sqrt{2} (a' b' -a b),\NL 
\bar{\mathcal{T}}_{25} =  a' \deet 
- a' b^2 + d' a^2 - 2 a b b', \NL[2mm]
\bar{\mathcal{T}}_{33}=d^2 + d'^2,\quad\bar{\mathcal{T}}_{34}=\sqrt{2} (b d + b' d'),\NL
\bar{\mathcal{T}}_{35}= d' \deet +  a' d^2 - d' b^2 + 2 b d b',\NL[2mm]
\bar{\mathcal{T}}_{44} = ad + b^2 + 2 b'^2,\quad 
\bar{\mathcal{T}}_{45} = \sqrt{2}\begin{array}[t]{l}\Big(b' \deet\\ 
+ a' b d  - d' a b  + b' ( a d + b^2)\Big), \end{array} \NL[2mm]
\bar{\mathcal{T}}_{55}= \begin{array}[t]{l}\deet^2 + a'^2 d^2 + d'^2 a^2\\
+ 4 b'^2 a d -2 a' d' b^2 + 4 a' b' b d - 4 b' d' a b  \end{array}\NL
\label{T5}
\eea    
and
\vspace{3mm}
\bea
\bar{\mathcal{T}}_{\pm 11} = a,\quad \bar{\mathcal{T}}_{\pm 12} = -b,\quad 
\bar{\mathcal{T}}_{\pm 13} = a d' \mp b b',\quad \bar{\mathcal{T}}_{\pm 14} = \mp a b' - b a',\NL
\bar{\mathcal{T}}_{\pm 22} = d,\quad \bar{\mathcal{T}}_{\pm 23} = \pm d b' - b d', \quad 
\bar{\mathcal{T}}_{\pm 24} = a' d \pm b b', \NL
\bar{\mathcal{T}}_{\pm 33} = a d^2 \mp 2 b b' d'^2 + d b'^2,\quad 
\bar{\mathcal{T}}_{\pm 34} = \mp a b' d' \pm d a' b' - b \deet, \NL
\bar{\mathcal{T}}_{\pm 44} = a b'^2 + d a'^2 \pm 2 b a' b'.
\label{T4pm}
\eea

In terms of the parameters $\Phi$ and $\Psi$ the quantities
$v$,  $a$, $d$, $b$ and $a'$, $d'$, $b'$ are given by:\\

\begin{equation*}
v = \frac{16 \,\left( 1-{{\it c_q}}^{2}{\Psi}^{2} \right)  \left( 1+{\Phi}^{4}
 \right) ^{2}-4\,{{\it s_q}}^{2}{\Phi}^{2}\Psi\, \left( 1+{\Phi}^{4}
 \right) +4\, \left( {\Psi}^{2}-{{\it c_q}}^{2} \right) {\Phi}^{4} }
   {\left( 1-{\Phi}^{2} \right) ^{4} \left( 1-\Psi^2 \right)^2 }\;;
\end{equation*}
\vspace{3mm}
\begin{equation*}
a = 2 c_q\; (c_q + 1) \Phi^2 (1 - \Phi^4) (1 - \Psi)^2 (1 - \Psi^2)/den\;,
\end{equation*}

\begin{equation*}
d = - a\big|_{q \to \pi + q}\;,
\end{equation*}

\begin{equation*}
b = a\; \frac{s_q}{ c_q + 1} 
\end{equation*}

with

\begin{equation*}
den = - (1 + \Phi^4 - 4 \Phi^2 \Psi
 + \Psi^2 + \phi^4 \Psi^2)^2 + 4 c_q^2  (1 - \Phi^2 \Psi)^2 (\Phi^2 - \Psi^2)^2\;.
\end{equation*}

One sees immediately that $a d - b^2 = 0$. This identity has been used in establishing 
the matrices in equations (\ref{T5}) and (\ref{T4pm}).

\vspace{3mm}
\begin{eqnarray*}
a'& = & \frac{s_q}{den}\Big(\big((1 + \Phi^4) (1 + \Psi^2)
 - 4  \Phi^2 \Psi \big)^2 \NL 
&{}& - 4 c_q (1 - \Phi^2 \Psi) ( \Psi - \Phi^2) \big(\Phi^2 (1- \Psi)^2
+ c_q \Psi (1 - \Phi^4)\big)\Big)\;,   
\end{eqnarray*}

\vspace{-2mm}

\begin{equation*}
d'= - a'\big|_{q \to q +  \pi}\;, \NL
\end{equation*}

\begin{eqnarray}
b'= \frac{c_q}{den} (1-\Phi^2)^2 \Big(&{}&(1+ \Psi^2)^2 (1 + \Phi^4)
 + 2 \Phi^2\big( (1 - \Psi^2)^2
- 2 \Psi - 2 \Psi^3 )\NL
&{}& - 4 c_q^2 \Psi (\Psi - \Phi^2)(1 - \Psi \Phi^2)\Big).
\end{eqnarray}  %

With this information explicit expressions for the eigenvalues and the eigenvectors of 
the transfer matrices that are needed in the main part of the paper can be obtained.
We refrain from presenting the rather lengthy explicit expressions for the coefficients 
$\phi_1 \dots \phi_4$ of the eigenvector (\ref{Gphig}) but we note certain symmetry 
properties of these coefficients that will be used to simplify expressions to be derived 
in the next appendix. First we observe that 

\be
\phi_4 = \phi_0,
\label{phi04}  
\ee

$\!\!\!\!$ and furthermore we find the following symmetries of $\phi_1$, $\phi_3$ and $\phi_3$ as 
functions of $q$: 

\be
\begin{array}{c|c|c|c|} &\phi_1\to&\phi_2\to&\phi_3\to \\ \hline
q\to-q&-\phi_1&-\phi_2&\phi_3\\
q\to\pi+q&-\phi_2&-\phi_1&-\phi_3\\ \hline
\end{array}
\label{symphiq}
\ee

\section{Calculation of the elements of the Toeplitz determinant}

In this appendix we present the main steps of the calculations that lead to the result
(\ref{cotoep}) for the elements $a_j$ of the Toeplitz determinant (\ref{toepchi}) in 
the main text. We need to determine the expectation value (cf.~equation (\ref{conxy}))

\bea
G_{m,n}  & = &  \frac{1}{\mathcal{N}_g} \langle \phi_g | i b^y_m b^x_n |\phi_g \rangle\;,
	 \quad \mbox{where} \quad \mathcal{N}_g = \langle \phi_g |\phi_g \rangle \NL
	 & = &  \frac{1}{\mathcal{N}_g} \langle \phi_g | - b_m\;b_n - (b_m\;b_n)^{\dagger} 
+ b_m^{\dagger}\,b_n + ( b_m^{\dagger}\,b_n)^{\dagger}\;|\phi_g \rangle (1- \delta_{m,n}) \NL
& {} & + \left(\frac{1}{\mathcal{N}_g} \langle \phi_g| 2\, b_m^{\dagger} b_m |\phi_g \rangle 
-1 \right)\; \delta_{m,n}\,,     
\label{Gmn}
\eea
cf.~equation (\ref{cxcy}). Let us first consider the expectation value
\be
f_{m,n} = \frac{1}{\mathcal{N}_g} \langle \phi_g |b_m\;b_n |\phi_g \rangle.
\label{fmn}
\ee

With the representation

\be
b_m =\frac{1}{\sqrt{N}} \sum_{0 \leq q < \pi/2}\!\!\left[e^{i q m} 
b_{1q}+ e^{-i q m}b_{2q}
+(-)^m e^{i q m} b_{3q}+(-)^m e^{-i q m} b_{4q} \right]
\label{redc}
\ee

of the annihilation operator at site $m$ by the momentum space operators $b_{1q}\;\dots b_{4q}$  
introduced in the main text (see paragraph below equation (\ref{fevg})) we find after 
a straight-forward 
evaluation of expectation values of the type  
$\;_q\langle\phi_g|b_{\mu q}\;b_{\nu q}|\phi_g \rangle_q$,\; ($\mu,\;\nu =1, \dots , 4$)     

\bea
f_{m,n} & = &\!\!\frac{1}{N} \,\sum_{0 \leq q < \pi/2}\frac{1}{\mathcal{N}_{gq}} 
\Big[\!\!-2\sin(q[m-n])\; \big(1- (-)^{m+n} \big)\; \phi_0 \big(\phi_1 + \phi_2) \NL
& {} & - 4 \cos(q[m-n])\; \big((-)^m - (-)^n \big)\; \frac{\phi_3}{\sqrt{2}} 
\phi_0 \Big]\;.
\label{prelfmn}
\eea 

Here
\be
\mathcal{N}_{gq} =\;_q\langle\phi_g|\phi_g \rangle_q = 2\phi_0^2 + \phi_1^2 + \phi_2^2 + \phi_3^2. 
\label{prelnorm}
\ee
(The momentum label $q$ has been suppressed on the r.h.s.) In  equations (\ref{prelfmn}) and 
(\ref{prelnorm}) use has been made of the identity (\ref{phi04}). 
Note that $f_{m,n}$ is real and 
hence 
\begin{equation*}
\frac{1}{\mathcal{N}_g} \langle \phi_g |b_n^{\dagger}\;b^{\dagger}_m |\phi_g \rangle
= \frac{1}{\mathcal{N}_g} \langle \phi_g |b_m\;b_n |\phi_g \rangle = f_{m,n}\;.
\end{equation*}
Similarly,
\bea
g_{m,n} & = & \frac{1}{\mathcal{N}_g} \langle \phi_g |b_m^{\dagger}\;b_n |\phi_g \rangle
\label{gmn}\\
& = & \frac{1}{N} \,\sum_{0 \leq q < \pi/2}\frac{1}{\mathcal{N}_{gq}} \Big[ 2 \cos(q[m-n]) 
\;\left\{ \phi_0^2 + \phi_1^2 + \frac{\phi_3^2}{2} \right. \NL  
&{}&\left. + (-)^{m+n}\;(\phi_0^2 + \phi_2^2+ \frac{\phi_3^2}{2}) \right\}\NL
& {} & - 2\sin(q[m-n])\;\frac{\phi_3}{\sqrt{2}}\;[(-)^m - (-)^n]\;(\phi_1 - \phi_2) \Big]\;.     
\label{prelgmn}
\eea

Obviously $g_{m,n}$ is real too, and with equation (\ref{prelnorm}) it follows immediately that
$g_{m,m} = 1/2$, so that the last term in the expression (\ref{Gmn}) vanishes. 
On account of the symmetries (\ref{symphiq}), the sums over  $q$ in Eqs. (\ref{prelfmn}), 
(\ref{prelgmn}) can be extended to the full 
Brillouin zone $-\pi \leq q < \pi$,
\begin{equation*}
\frac{1}{N}\sum_{0 \leq q < \pi/2} \cdots\; \longrightarrow\;
 \frac{1}{4 N}\sum_{-\pi \leq q < \pi} \cdots\;.  
\end{equation*} 
Then,

\bea
G_{m,n} & = & 2\;( - f_{m,n}\;+\; g_{m,n}) (1 - \delta_{m,n})\NL
& = & \frac{1}{2}\big[1 - (-)^{n-m} \big]\;\gamma_{n-m}
 + \frac{1}{2}\big[(-)^{m} - (-)^n \big]\;\gamma_{n-m}'\NL
\mbox{with}\NL
\gamma_l & = & \frac{4}{N}\!\!\sum_{-\pi \leq q < \pi}\; \frac{1}{\mathcal{N}_{gq}}\;
\frac{1}{4} (\phi_1 + \phi_2)\;\big[
(\phi_1 - \phi_2) \cos(ql)- \;2 \phi_0\; \sin(ql)\big] \NL
\mbox{and}\NL
\gamma_l' & = & \frac{4}{N}\!\sum_{-\pi \leq q < \pi}\;\frac{1}{\mathcal{N}_{gq}}\; 
\frac{\phi_3}{\sqrt{2}}\; \big[ \phi_0\;
\cos(ql) + \frac{1}{2} (\phi_1 - \phi_2)\; \sin(ql)\big].
\label{1Gmn}
\eea

For the elements of the determinant (\ref{detchi}) this yields

\bea
G_{2m,2m'} & = & 0\;,\qquad G_{2m+1,2m'+1} = 0 \NL
G_{2m-1,2m'} & = & \gamma_{2(m'-m)+1} - \gamma_{2(m'-m)+1}' \equiv a_{m'-m}' \NL
G_{2m,2m'+1} & = & \gamma_{2(m'-m)+1} + \gamma_{2(m'-m)+1}' \equiv a_{m'-m}\;.
\label{2Gmn}
\eea

Further simplifications of these expressions are possible on account of the following 
properties of the coefficients $\phi_1$, $\phi_2$ and $\phi_3$ that were found by analysing 
the explicit expressions for these coefficients with computer algebra:
They have a common denominator $\bar{D}_0(q)$ which is invariant under the operations $q \to -q$ 
and $q \to \pi+q$,

\be
\phi_{\nu}(q) = \frac{\bar{D}_{\nu}(q)}{\bar{D}_0(q)}, \qquad \nu\;=\;1, 2, 3.
\label{phis}
\ee

Only two of the four coefficients $\bar{D}_0$, $\bar{D}_1$, $\bar{D}_2$ and $\bar{D}_3$ 
are independent. The relations between them are

\bea 
-\frac{1}{\sqrt{2}} c_q \bar{D}_3(q) - \frac{1}{2} s_q  \big[\bar{D}_1(q) + \bar{D}_2(q) \big] 
& = & \bar{D}_0(q)\NL[2mm]
 \mbox{and}\NL[2mm]  
 -\frac{1}{\sqrt{2}} s_q \bar{D}_3(q) + \frac{1}{2} c_q \big[\bar{D}_1(q) + \bar{D}_2(q) \big]
 & = &  
\frac{1}{2} (\bar{D}_1(q) -  \bar{D}_2(q)).
\label{relD}
\eea

Using these relations one finds 
\be
a_j' = \delta_{j,0}
\label{apj}
\ee
and, introducing $p = 2q$ as the momentum variable  and denoting 
$\bar{D}_{\nu}(\frac{p}{2}) \equiv D_{\nu}$,
\bea
a_j & = & \frac{1}{N} \sum_{-\pi \leq p < \pi} 
\Big\{ \left[\;D_0^2 - \frac{1}{4}\big(D_1 - D_2 \big)^2\right]\;\cos(p[j+1])\NL
& {} & \quad + D_0\, (D_1 - D_2)\;\sin(p[j+1])\Big\} \frac{1}{D_0^2 +
\frac{1}{4}\big(D_1 - D_2 \big)^2} .
\label{ajs}
\eea

In the continuum limit, $N \to \infty $, and with $z = e^{ip}$ as the integration variable, this 
leads to the expression (\ref{cotoep}) for $a_j$.

\section{Perturbation expansion in $J'/T$}

As we have mentioned  in the Introduction, for $J > J'$ the chain spins of the AIK model are 
ideally ordered antiferrromagnetically in the ground state, and as a consequence the middle spins 
are completely decoupled from the chain spins. For weak interchain coupling $J' \ll J$ the order  
of the spins along the chains will prevail over a long distance for low but finite temperatures 
$T \ll J$, and the coupling between the chains will remain small as long as $J' \ll T$. In the 
sequel we will calculate the correlation function between the middle spins of row $l$ of 
the kagome lattice, $\chi(2m+1, 2n+1) = \langle \;\sigma_{2m+1,l}\; \sigma_{2n+1,l}\;\rangle$,    
perturbatively with respect to $J'$. 
(In $\sigma_{j,l}$ the second subscript $l$ denotes the row $l$ of the 
 lattice.) The hamiltonian of the AIK model couples the middle spins of row $l$ to 
chain spins of the rows 
$l-1$ and $l+1$,

\be
 J' \; \sigma_{2m+1,l} \cdot
 \left[\sigma_{2m+1,l-1} + \sigma_{2m,l-1} + \sigma_{2m+1,l+1} + \sigma_{2m,l+1}\right]\;.
\label{ham}
\ee

In second order in $J'/T$ one obtains for $\chi$:

\bea \arraycolsep0pt
&&\chi(2m+1, 2n+1) \NL &&= 2\;(\frac{J'}{T})^2 \langle 
 \left[\sigma_{2m+1,l+1} + \sigma_{2m,1}\right]
 \left[\sigma_{2n+1,l+1} + \sigma_{2n,1}\right]\rangle_{J'=0} 
 + \mO\left((\frac{J'}{T})^3\right) \NL
&&= 2\;(\frac{J'}{T})^2\; \Big[ 2\;\chi_{\mbox{\tiny 1d-Ising}}(2n-2m)
 + \chi_{\mbox{\tiny 1d-Ising}}(2n-2m+1) \NL
&&\qquad\qquad\;\; + \chi_{\mbox{\tiny 1d-Ising}}(2n-2m-1) \Big]
 + \mO\left((\frac{J'}{T})^3\right) 
\eea

Here $\chi_{\mbox{\tiny 1d-Ising}}(n)$ is the correlation function of the one-dimensional Ising 
chain,

\be
 \chi_{\mbox{\tiny 1d-Ising}}(n) = (-)^n \left( \tanh(\frac{J}{T}) \right)^{|n|}\;.
\ee

Hence
\be
 \chi(2m+1, 2n+1) = -\;\frac{8 (J'/T)^2}{e^{4J/T}-1} \;
 \left( \tanh(\frac{J}{T}) \right)^{2|n-m|} + \mO\left((\frac{J'}{T})^3\right)\;.
\label{pertmidchi}
\ee

In the limit considered here, the correlation between any pair of  middle spins of the same row is 
negative for all distances. It is mediated by the interaction with the neighbouring spin 
chains.

\newpage
\section*{References}
\bibliography{frisikago.bbl}
%!%\input{frisikago.bbl}

\end{document}